\documentclass[%
 reprint,
 amsmath,amssymb,
aip,
]{revtex4-2}

\usepackage{graphicx}
\usepackage{dcolumn}
\usepackage{bm}

\usepackage{times}

\usepackage{nicefrac} 
\usepackage[hidelinks]{hyperref} 
\usepackage{graphicx}
\usepackage{color}
\usepackage{siunitx} 
\DeclareSIUnit\Angs{\angstrom}

\newcommand{\tise}{TiSe$_2$~}
\newcommand{\ttise}{TiSe$_2$}

\newcommand{\flux}{\,\nicefrac{$\mu J$}{$cm^2$}\,}
\newcommand{\fflux}{\,\nicefrac{$\mu J$}{$cm^2$}}

\begin{document}

\preprint{APS/123-QED}

\title{Revealing the order parameter dynamics of 1T-\tise following optical excitation}%
\author{Maximilian Huber}
\author{Yi Lin}%
\affiliation{%
 Materials Science Division, Lawrence Berkeley National Laboratory
Berkeley, CA 94720, USA
}%
\author{Nicholas Dale}
\affiliation{%
 Materials Science Division, Lawrence Berkeley National Laboratory
Berkeley, CA 94720, USA
}%

\affiliation{%
 Physics Department, University of California Berkeley, Berkeley, CA 94720, USA
}%


\author{Renee Sailus}
\author{Sefaattin Tongay}
\affiliation{
 Materials Science and Engineering Department, Arizona State University,
AZ 85281, USA
}%

\author{Robert A. Kaindl}

\affiliation{%
 Materials Science Division, Lawrence Berkeley National Laboratory
Berkeley, CA 94720, USA
}%
\affiliation{%
 Department of Physics and CXFEL Labs, Arizona State University, AZ 85287, USA \\
 * Email: alanzara@lbl.gov
}%
\author{Alessandra Lanzara*} 

\affiliation{%
 Materials Science Division, Lawrence Berkeley National Laboratory
Berkeley, CA 94720, USA
}%

\affiliation{%
 Physics Department, University of California Berkeley, Berkeley, CA 94720, USA
}%

\date{\today}

\begin{abstract}
The formation of a charge density wave state is characterized by an order parameter. The way it is established provides unique information on both the role that correlation plays in driving the charge density wave formation and the mechanism behind its formation. Here we use time and angle resolved photoelectron spectroscopy to optically perturb the charge-density phase in 1T-\tise and follow the recovery of its order parameter as a function of energy, momentum and excitation density. Our results reveal that two distinct orders contribute to the gap formation, a CDW order and pseudogap-like order, manifested by an overall robustness  to optical excitation. A detailed analysis of the magnitude of the the gap  as a function of excitation density and delay time reveals the excitonic long-range nature of the CDW gap and the short-range Jahn-Teller character of the pseudogap order. In contrast to the gap, the intensity of the folded Se$_{4p}$* band can only give access to the excitonic order. These results provide new information into the the long standing debate on the origin of the gap in \tise and place it in the same context of other quantum materials where a pseudogap phase appears to be a precursor of  long-range order. 
\end{abstract}

\maketitle


Charge-density waves (CDW) are states of broken symmetry characterized by a complex order parameter defined by both phase and amplitude, where the latter is proportional to the energy gap\,\cite{Gruener}. In the simplest approximation they can be described by a  mean-field theory of a one-dimensional electron–lattice  system in the weak-coupling limit\,\cite{Rossnagel2011, Gruener} which leads to a large coherence length. Note that the term coherence length, which will be used in this work analogously to classical CDW literature\,\cite{McMillan, Gruener}, corresponds to the spatial dimension of electron-electron or electron-hole pairs\,\cite{Gruener}. In contrast to weak coupling, within the strong coupling limit the coherence length becomes short and the system can be described best within a local bonding picture. As a consequence, in the strong coupling limit only the long-range order of the CDW is lost above the transition temperature but fluctuating short-range order remains\,\cite{McMillan, Rossnagel2011}. In other words, amplitude and phase coherence become disentangled and a gap in the electronic spectra, a pseudogap, appears well before the long-range charge density wave order sets in. In this pseudogap phase short-ranged CDW fluctuations with a well defined amplitude are still present. We note that this phase is different from the high symmetry phase predicted to exist at high temperature where the system is semi-metallic with a negative gap\,\cite{Hellgren2017, Lian2020, Bianco2015}.
This behavior bears analogies to the case of unconventional superconductors\,\cite{Kanigel2008, Ding1996}. \\
Among the various CDW materials, 1T-\tise holds a special place given the presence of a) a broad variety of many body interactions, such as electron-electron, electron-hole and electron-lattice interactions, leading to excitonic condensation\,\cite{Kogar2017, Monney2009} and phonon softening\,\cite{Weber2011, Otto2021} b) of superconductivity induced via pressure\,\cite{Kusmartseva2009, Joe2014} or doping\,\cite{Ramirez2006},  both of which lead to a breaking of the CDW long-range order; and c) CDW fluctuations above the transition temperature, suggestive of a strong coupling regime and short coherence length\,\cite{Weber2011, Rossnagel2011, Hildebrand2016}. \\
The key to study this intriguing interplay of long- and short-range order and to understand the emergence of such correlated behavior is the order parameter. Indeed, having direct access to the quenching and re-formation dynamics of the order parameter during a photoinduced phase transition retains unique information on the mechanism behind the establishment of the new order. In this regard, time- and angle-resolved photoemission spectroscopy (tr-ARPES) yields a powerful way to access the formation and dynamics of the gap, given that it is the only technique that can directly monitor the onset of the order parameter with simultaneous momentum and time resolution. While signatures of the gap in the optical conductivity\,\cite{Porer2014a} as well as the dynamics of the valence band at the $\Gamma$ point\,\cite{Hedayat2019a, Duan2021}  have been studied before, in this study we directly follow the actual gap dynamics while also providing a direct comparison with all other relevant quantities, namely excited carriers and backfolded Se$_{4p}$* band. We find that following photoexcitation, the gap undergoes only a partial quenching, up to 30$\%$, of its equilibrium value ($\sim$130\,meV). This points to the existence of a pseudogap and to the presence of multiple coexisting mechanisms contributing to the CDW formation in \ttise.  Finally we reveal that the folded Se$_{4p}$* is mostly connected to only one of the components cautioning the common assumption that spectral weight dynamics is directly related to order parameter dynamics.



\begin{figure} [h]
	\centering
	\includegraphics[width=0.5\textwidth]{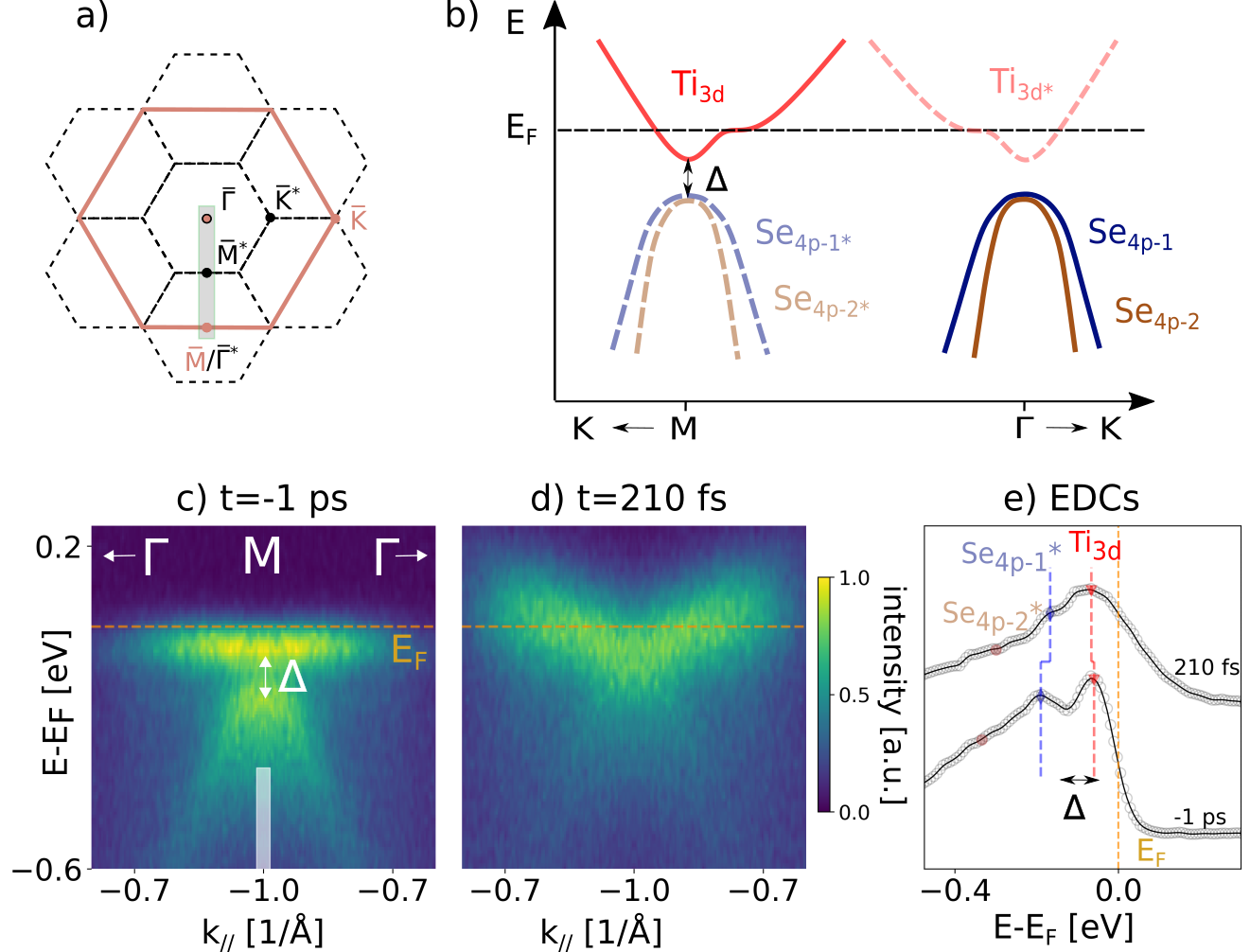}
	\caption{ a) First BZ in the high symmetry (orange) and CDW phase (black); the grey bar illustrates the cuts measured in the ARPES spectra in $\Gamma$-$M$-$\Gamma$ direction. b) Schematic bandstructure of the CDW state (adapted from\,\cite{Lian2020}).  c) and d) ARPES spectra of the $M$-point in equilibrium\,(c) and after excitation with 780\,nm pump pulses at 80\,\flux fluence (d). For clarity the spectra are mirrored at the $M$ point. The Fermi level is indicated by the dashed orange line. e) EDCs taken at $M$ corresponding to the spectra in panel\,c and\,d. The region in momentum space over which the EDCs were integrated is shown by the white box in panel\,c. Markers indicate the fitted peak positions of the Ti$_{3d}$ conduction band (red triangle) as well as the Se$_{4p-1}$ (blue diamonds) and Se$_{4p-2}$ valence bands (brown circles). Circles represents raw data and solid black lines represent the smoothed raw data using the Gaussian method (10\,meV window).}
	\label{fig:fig1}
	\end{figure}

\autoref{fig:fig1} shows the dynamics of the electronic band structure along the high symmetry direction $\Gamma$-$M$-$\Gamma$ for a temperature of 80\,K, i.e.\ below the CDW transition. 

Data are taken for a probe energy of 22.3\,eV with pump excitation of 780\,nm ($\sim$1.6\,eV) and a fluence of 80\,\fflux. The repetition rate is 25\,KHz and the overall temporal resolution is $\sim$ 65\,fs. More details of the XUV-trARPES setup can be found in the Methods section of the Supplementary Material as well as in\,\cite{Buss2019, Wang2015}. \\
Note that the probe energy of 22.3\,eV allows access to states close to $A$ and $L$ point within the bulk Brillouin zone notation\,\cite{Watson2019}. However, for consistency with previous trARPES work\,\cite{Mathias2016, Monney2016}, we will use the surface BZ notation and refer to these points as the $\Gamma$ and $M$ point\,\cite{Monney2010a}. Panel\,a shows the 2D Brillouin zone (BZ) in the high symmetry (orange) and CDW (black) state. When going through the phase transition, the unit cell size doubles from a (1$\times$1$\times$1) structure at room temperature to a (2$\times$2$\times$2) superstructure below $\sim$200\,K\,\cite{DiSalvo1976}. Panel\,b displays the schematic band structure\,\cite{Lian2020} in the CDW phase along the high symmetry directions.  It shows the bands from Ti$_{3d}$ to Se$_{4p}$ orbitals. As widely discussed in the literature\,\cite{Rossnagel2002, Rossnagel2010, Rossnagel2011, Rohwer2011, Kidd2002}, the fingerprints of the CDW state are the backfolding of the spin-orbit split Se$_{4p}$ bands at $\Gamma$ (Se$_{4p-1}$ and Se$_{4p-2}$, see solid line in panel b) onto the $M$ point (Se$_{4p-1}$* and Se$_{4p-2}$*, see dashed line in panel b) and the opening of a gap. Panel\,c shows the measured equilibrium band structure measured along the $\Gamma$-$M$-$\Gamma$ direction (location of the cut in the Brillouin zone is indicated by the grey box in panel a), where both the backfolded Se$_{4p-1}$* and Se$_{4p-2}$* valence band are visible. Additionally we can resolve the bottom of the Ti$_{3d}$ conduction band and hence access the gap formed between the lowest Ti$_{3d}$ and the highest Se$_{4p}$* band. Panel d\,shows the evolution of these bands following pump excitation, revealing three key features. First, the the Ti$_{3d}$ parabolic band at $M$ point is occupied by hot electrons. Moreover, we observe a fast reduction of the gap size between the Ti$_{3d}$ and Se$_{4p-1}$* states. Finally the intensity of the Se$_{4p-1}$* backfolded band is reduced along with the the quenching of the gap.
\\ 
The combination of, compared to previous XUV-trARPES work\,\cite{Rohwer2011, Mathias2016, Hellmann2012, Rohde$_2$013, Rohde$_2$014}, higher energy resolution ($<80\,$meV) and access to lower pump fluences of this study allows us for the first time to access the gap dynamics following optical excitation. In panel\,e  of \autoref{fig:fig1}, we present raw energy distribution curves (EDCs) at $M$ point measured along  $\Gamma$-$M$-$\Gamma$- direction (the region in momentum space over which the EDCs were integrated is indicated by the white box in panel\,c) corresponding to the two spectra shown in panel\,c and\,d. In equilibrium we clearly resolve two peaks in the raw spectra corresponding to the top of Se$_{4p}$* valence band and the bottom of Ti$_{3d}$ conduction bands. The distance between the two peaks defines the energy gap. The lower, spin split Se$_{4p-2}$* band is also resolved in the raw spectra as a weak shoulder. After 210\,fs the three bands are still visible, however excited carriers and the melting of the Se$_{4p}$* band make the gap, despite being still open, less well defined. To extract the gap size and its time evolution, the EDCs spectra  were fitted with a function using three peaks of Voigt lineshape to account for the Ti$_{3d}$ as well as the two spin splitted Se$_{4p}$* bands. Details of the fitting procedure are given in Supplementary Note\,3. The fitted peak positions are marked in panel\,e, where red, blue and brown symbols correspond, respectively, to the Ti$_{3d}$, Se$_{4p-1}$* and Se$_{4p-2}$* band.


\begin{figure*} 
	\centering
	\includegraphics[width=0.8\textwidth]{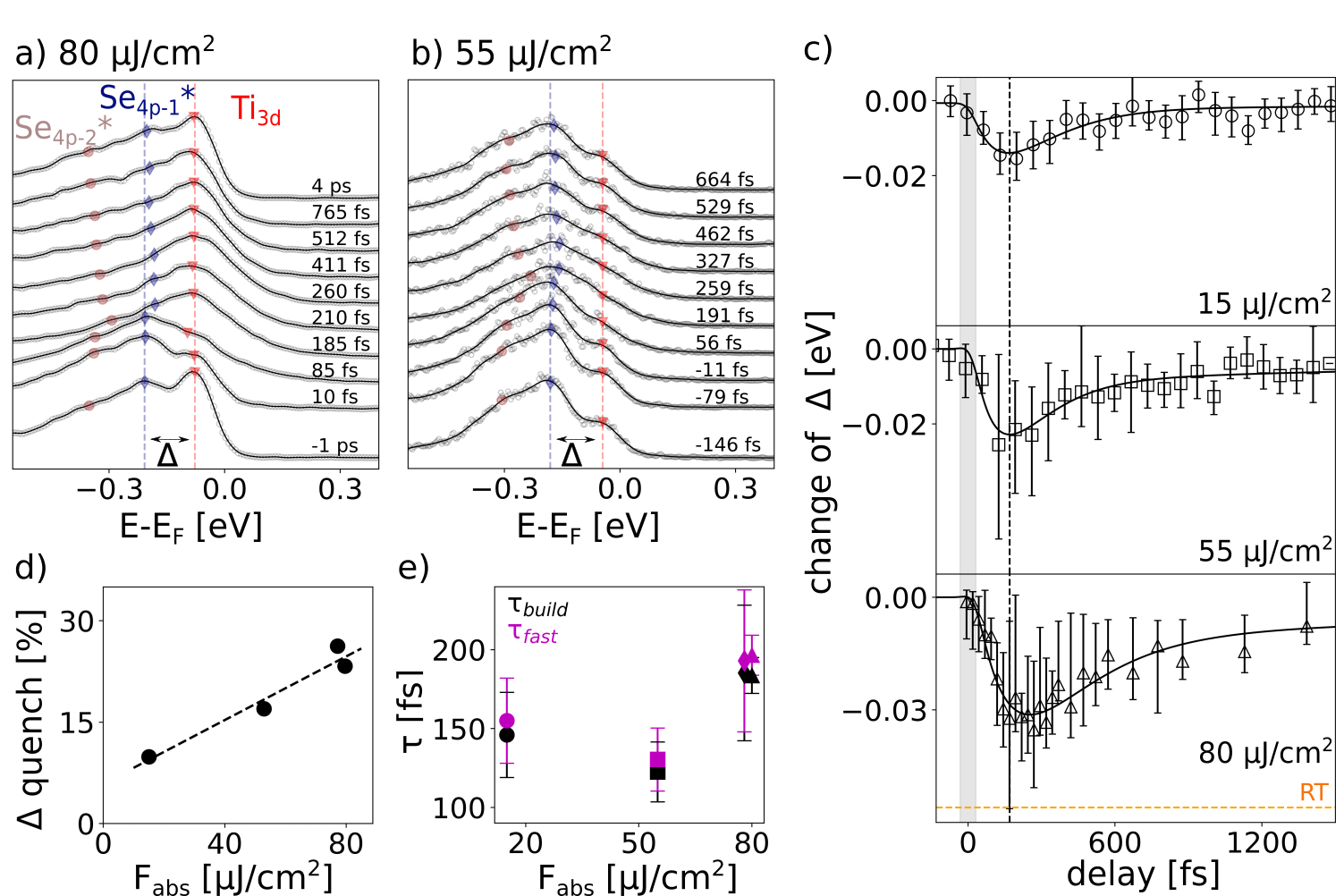} 
	\caption{Gap Dynamics at different excitation density. a-b)  Selected EDCs taken at $M$ point with markers indicating the fitted peak positions of the Ti$_{3d}$ conduction band (red triangle) and the Se$_{4p-1}$ (blue diamonds) and Se$_{4p-2}$ valence bands (brown circles) after excitation with 80 (panel a) and 55\,\flux (panel b), respectively. Circles represents raw data and solid black lines represent the smoothed raw data using the Gaussian method (10\,meV window).  (c) Fitted change of gap size over delay after excitation with 15,55 and 80\,\fflux. Solid black lines represent fits to the data. The dotted orange line  represents the gap size extracted by equilibrium ARPES at room temperature\,\cite{Watson2019}.  d) Maximum gap quench (with respect to the equilibrium value) in dependence of absorbed fluence F\textsubscript{abs}. Dashed line is a linear fit of the data points.  e) Fitting parameters $\tau\textsubscript{build}$ and $\tau\textsubscript{fast}$   in dependence of the absorbed fluence. }
	\label{fig:fig2}
	\end{figure*}

To study in detail the response of the gap to optical excitation, in \autoref{fig:fig2} we present the fluence dependence of the gap and its dynamics. Panels\,a and\,b show EDCs taken at different delay times at $M$ point after excitation with 80 (panel a) and 55\,\flux (panel b) to show the temporal evolution of the gap. Note that the slightly different shape of the EDCs in equilibrium is due to the fact that a different batch of samples was used (see Methods section in the Supplementary Material). While in panel\,b the gap remains clearly visible over all delay times, the pronounced melting of the backfolded Se$_{4p}$* band in panel\,a obscures its exact peak position for intermediate delay times ($\sim$ 250-550\,fs). Despite this difficulty, we can still extract the gap size with high precision as our XUV light source allowed us to simultaneously monitor the dynamics of the Se$_{4p}$ band at $\Gamma$ point (see Supplementary Note\,3). In panel\,c we plot the change of the gap size as a function of delay time for various fluences, referenced to the equilibrium gap size of 132$\pm 3$\,meV. For reference we also plot the gap size at room temperature (74\,meV\,\cite{Watson2019}) (dotted orange line). Additional gap data after excitation with a fluence of 78\,\flux is shown in Supplementary Note\,4. A remarkable fact is that the gap consistently only gets quenched about a fraction of its equilibrium value. This is further confirmed by the gap extracted from spectra taken along $K$-$M$-$K$ direction (see Supplementary Note\,4).

Panel\,d shows the fluence dependence of the relative gap quench. The data can be appropriately described by a linear fit. However we note that this relationship only holds for the intermediate fluence regime studied in this work. Further studies are needed to elucidate the behavior at very low fluences. For the highest fluences studied, the gap gets quenched only about $<$30$\%$ of its equilibrium value. This is a surprising result, as for these fluences, which are well above the reported critical fluence for a complete destruction of the excitonic condensate\,\cite{Porer2014a} of 40\fflux, X-ray diffraction experiments\,\cite{Burian, Mohr-Vorobeva2011, Cheng2022} found a considerable, almost complete, suppression of the structural long-range order. Moreover in our experiment the electronic temperature reaches about 1000\,K (as extracted from the Fermi level broadening), which is considerably over the equilibrium transition temperature of 200\,K.  This robustness of the gap to optical excitation contrasts with various theoretical calculations, which consistently predict the high symmetry phase of \tise to be semi-metallic\,\cite{Lian2020, Hellgren2017, Bianco2015, Singh2017}. At the same time this robustness is analogous to ultrafast THz\,\cite{Porer2014a} and equilibrium ARPES studies\,\cite{Watson2019, Rossnagel2002, Monney2010a}, where the gap remains open at high excitation densities and temperatures, and is reminiscent of a pseudogap behavior where, even at very high excitation fluences (or temperatures above the transition temperature) the gap is only partially affected\,\cite{Smallwood2014, Kanigel2008, Ding1996}. Finally this observation clearly suggests that the CDW phase in \tise differs from the behavior of other CDW systems\,\cite{Schmitt2008, Hellmann2012, Perfetti2006, Petersen2011} and superconductors\,\cite{Smallwood2014, Zhang2014, Smallwood2015} where the order parameter can be optically fully quenched.

Besides the extent of the gap quench additional information on the gap formation mechanism can be obtained by studying the  dynamical behavior of the gap. To analyse the dynamics, we fit the data in panel\,c with the sum of three exponential functions,  where one exponential accounts for the quench time, one for a fast and one for a slow recovery component (see Supplementary Note\,5). In the following, we refer to the time where the photoinduced change reaches its maximum, as the build up time $t_{build}$\cite{Rohwer2011, Zhang2016}. The extracted fitting parameters, describing the characteristic timescales for the buildup time ($\tau\textsubscript{build}$) as well as the short ($\tau\textsubscript{fast}$) relaxation times are shown in \autoref{fig:fig2}e. At low fluences the gap dynamics is dominated by a fast femtosecond relaxation process and, as the excitation strength increases, a slower component becomes more dominant for the gap dynamics extracted along $\Gamma$-$M$-$\Gamma$ direction. As a consequence of this slow recovery dynamics, full reopening of the gap occurs on an extended multi picosecond timescale. Furthermore, there is overall only a weak fluence dependence for $\tau\textsubscript{build}$ and $\tau\textsubscript{fast}$, with a slight increase with increasing fluence. A similar behavior was observed studying the dynamics of only the Se$_{4p}$ valence band at $\Gamma$ point\,\cite{Hedayat2019a, Duan2021}. The partial quenching of the gap \autoref{fig:fig2} suggests the presence of two different processes governing the recovery of the CDW phase; the impeded recovery after strong excitation might be explained by a phonon bottleneck mechanism\,\cite{Hedayat2019a} or to the slow re-establishment of long range coherence because of the inhomogeneous pseudogap state, photoinduced defects\,\cite{Zong2019} and potentially a dimensional crossover\,\cite{Cheng2022}.

\begin{figure*} 
	\includegraphics[width=0.9\textwidth]{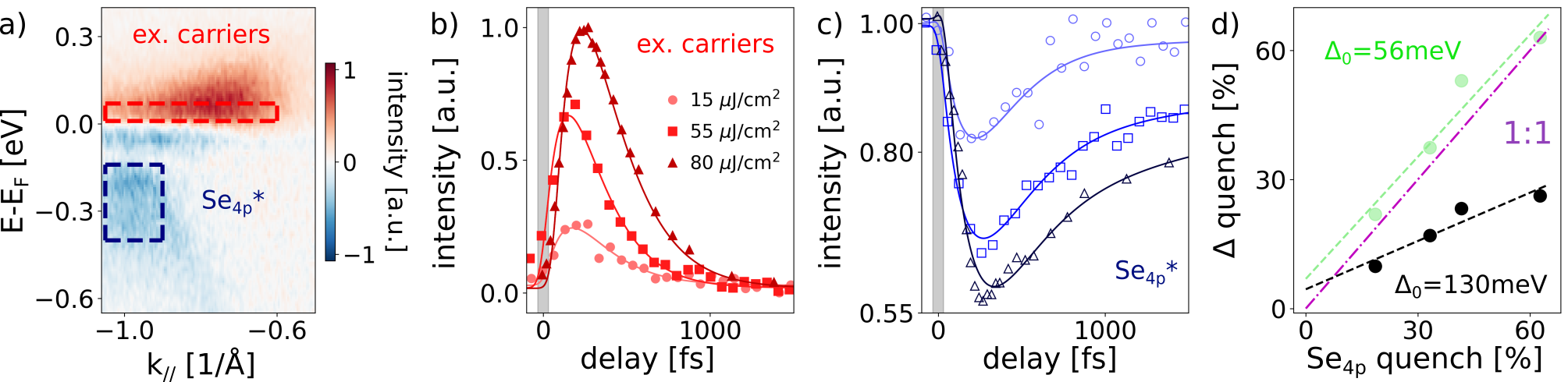}
	\caption{ a) Difference spectrum between a delay of -1\,ps and 185\,fs after excitation with 80\,\flux at 80\,K. b) Integrated intensity of excited carriers right above the Fermi level (integration region shown by red box in panel a)  after excitation with different fluences. Solid lines represent fits to the data points. c) Integrated intensity of Se$_{4p}$* band (integration region shown by blue box in panel a) after excitation with different fluences.  d) Maximum relative gap quench with respect to its equilibrium size in dependence of the maximum quench of the Se$_{4p}$* intensity. Green data points show the relative gap quench relative to a reference gap size $\Delta_0$ of 56\,meV.  Dashed lines are linear fits to the data points. Magenta line shows a 1:1 relationship for comparison.}
	\label{fig:fig3}
	\end{figure*}

To shed light into the gap formation mechanism, we compare the gap dynamics with the fluence dependent temporal behavior of the Se$_{4p}$* band and the excited carriers right above the Fermi level and extent the line of previous XUV measurements\,\cite{Rohwer2011, Mathias2016} into the low fluence regime. To extract the dynamics, we integrate the photoemission counts in the red box and the blue box of \autoref{fig:fig3}a as the intensity measure of the excited carriers and folded Se$_{4p}$* bands, respectively. A detailed discussion for the choice of this particular integration region of the excited carriers can be found in Supplementary Note\,7.
Panel\,b in \autoref{fig:fig3} shows the excited carrier dynamics at different excitation fluences and panel\,c the corresponding intensity curves of the Se$_{4p}$* band. All curves are normalized and show the relative change in spectral weight following optical excitation (see Supplementary Note\,1 for the normalization method and details on the background subtraction). To quantitatively analyze these intensity dynamics, we apply the same fitting method as we described for analyzing the gap dynamics in \autoref{fig:fig2}. \\
Notably, the higher the absorbed fluence, the more the Se$_{4p}$* band gets quenched (panel c) and an the more free carriers get excited (panel b). Interestingly, for the same fluence the Se$_{4p}$* band gets clearly more quenched than the gap (\autoref{fig:fig2}d). For a detailed quantitative comparison, in panel\,d we put the relative gap quench in direct relation to the Se$_{4p}$* quench.  It shows that consistently the extent to which the gap gets quenched is considerably lower than how much the Se$_{4p}$* intensity gets quenched. Interestingly, when assuming that only the part of the gap that can get quenched by temperature, i.e.\ $\Delta_{0} \sim$130\,meV-74\,meV=56\,meV, with 74\,meV being the gap size at room temperature\,\cite{Watson2019, Monney2010a}, is the "real" equilibrium gap, then the new relative quenches (shown in green) of the gap size is considerably closer to a 1:1 relationship (magenta line)  with the relative quenches of the Se$_{4p}$* intensity. This is a strong indication that the majority of the spectral weight of the Se$_{4p}$* band is connected to only a part of the gap (the part that gets more easily quenched) and thus have the same origin, whereas another contribution to the gap exists that is less pronounced in the Se$_{4p}$* replica band intensity. We note that within the studied fluence range we do not observe a complete quench of the Se$_{4p}$* band; the highest quench observed is $\sim 60 \%$ at 80\,\flux, which is above the reported threshold fluence of 40\flux for a complete destruction of the excitonic condensate\,\cite{Porer2014a}. On the one hand a precise comparison of absorbed fluences is difficult, hence it might be possible that even our highest fluence studied is still below the threshold fluence, on the other hand the residual intensity might be caused by the pseudogap-like contribution, which is less pronounced than in the gap.



\begin{figure} 
	\includegraphics[width=0.5\textwidth]{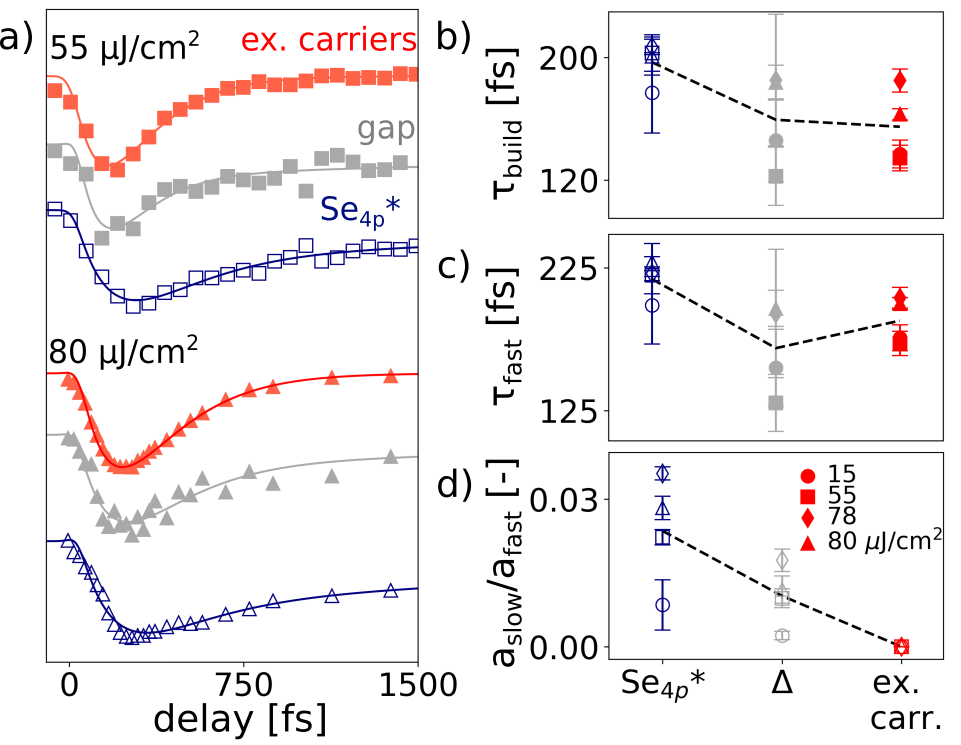}
	\caption{ a) Comparison of the dynamics of Se$_{4p}$* intensity (blue empty markers), excited carriers (red solid markers) and gap (grey solid markers) for two selected fluences. For clarity the excited carrier dynamics is flipped upside down. b-d) Extracted fitting parameters $\tau\textsubscript{build}$ (b), $\tau\textsubscript{fast}$ (c) and $a\textsubscript{slow}/a\textsubscript{fast}$ (d) of excited carriers, Se$_{4p}$* band intensity and gap for various fluences. Black dashed lines connect the values averaged over all fluences and serve as guide to the eye.}
	\label{fig:fig4}
	\end{figure}

For a further analysis we directly compare the dynamics of Se$_{4p}$* band, gap and excited carriers for two selected fluences in panel\,a  of \autoref{fig:fig4}. While an excellent correlation between excited carriers and gap dynamics up to $\sim$500\,fs is observed for all fluences studied (see Supplementary Note\,6), the Se$_{4p}$* photoinduced intensity changes seem to indicate a much slower initial recovery dynamics. To quantify these findings, in panel\,b-d we show the fitted parameters for both gap, excited carriers and Se$_{4p}$* intensity for different fluences. Similarly to the gap (\autoref{fig:fig2}c) two different recovery dynamics are identified for each of these curves, leading to a short ($\tau\textsubscript{fast}$) and a long ($\tau\textsubscript{slow}$) relaxation timescale, in addition to the buildup time ($\tau\textsubscript{build}$).  As qualitatively seen in panel\,a, the data in panel\,b and\,c reveal overall similar timescales for the gap and the excited carriers for both the buildup time $\tau\textsubscript{build}$ and the initial recovery $\tau\textsubscript{fast}$. The similarity in buildup time $\tau\textsubscript{build}$ between gap and hot electrons is in favor of an CDW phase that is directly driven by electronic effects. Indeed, as most electrons get photoexcited to $\sim$0.6\,eV above E$_F$ (see Supplementary Note\,6), carriers accumulating right above the Fermi level are mostly electrons that have scattered down from higher energy levels or electrons that got created in a carrier multiplication process\,\cite{Mathias2016,Saha2021}, indicating a strong and direct connection between the decreasing of the order parameter and the scattering rate of excited carriers. In contrast, if the quench of the gap was driven by the lattice, one would expect a delay in the response of the gap compared to the build up time of the excited carriers. Accordingly, a recent ultrafast electron diffuse scattering (UEDS) study found that the free carriers created by optical excitation are not strongly coupled to the phonon modes in \ttise\,\cite{Otto2021}. \\
Compared to gap and excited carriers the build up time and especially the initial recovery dynamics of the Se$_{4p}$* intensity is by inspection of panel\,a seemingly much slower. This is reflected in the values of $\tau\textsubscript{build}$ and $\tau\textsubscript{fast}$ in panel\,b and\,c. Furthermore, the three quantities differ by how pronounced the long term component with respect to the fast component is, i.e.\ the value of  $a\textsubscript{slow}$/$a\textsubscript{fast}$, as plotted in panel\,d. While the excited carriers do not show a slow component at all and go back to their equilibrium value within $\sim$1\,ps, $\tau\textsubscript{slow}$ clearly influences the gap dynamics and is the most pronounced in the Se$_{4p}$* band dynamics. Note that within our fitting model both $\tau\textsubscript{fast}$ and $a\textsubscript{slow}$/$a\textsubscript{fast}$ modulate the initial recovery between $\sim$150-600\,fs, the range in which the differences between gap and Se$_{4p}$* intensity dynamics are the clearest, and one cannot unambiguously determine if the fast or slow recovery contribution is more responsible for the observed differences. Nevertheless, while \autoref{fig:fig3}d implies that the part of the gap that is more easily quenchable and the majority of the Se$_{4p}$* intensity are predominantly of the same origin, our detailed analysis hints towards subtle differences in the dynamics between both quantities. Unfortunately, to the best of our knowledge their interplay has not been studied yet, neither experimentally nor theoretically, and therefore further investigation is necessary to further clarify the exact relationship between these two spectral fingerprints of the CDW state and to check if the observed differences are a real effect. \\ \\



The ability to directly access the gap and its dynamic has allowed us to reveal an intriguing unconventional CDW state in \ttise, where two different orders contribute to the gap, one which is easily quenched and a one more robust to photoexcitation. In this regard the sum of our experimental observations allows us to propose a coherent microscopic interpretation of these two order parameters. 
We first focus on the origin of the part that gets quenched, which we argue to display mostly excitonic order due to multiple reasons. First, the correlation of the buildup time $\tau\textsubscript{build}$ (as shown in \autoref{fig:fig4}) between gap and hot electrons is strongly in favor of an electronically driven CDW state. As argued above, the similarities in build up time indicate a strong and direct connection between the decrease of the order parameter and the scattering rate of excited carriers. This can be interpreted as the destruction of an excitonic insulator state, where the driving force is the transfer of kinetic energy into the excitonic condensate\,\cite{GoleZ2016, Saha2021}. Second, we established a direct connection between the quenched part of the gap and the backfolded Se$_{4p}$* band intensity (\autoref{fig:fig3}d). Previous trARPES studies focusing on the latter found that it displays behavior characteristic for the excitonic insulator, namely a strongly fluence dependent melting time\,\cite{Rohwer2011}. Moreover, it was argued that the high intensity of the backfolded band itself in equilibrium can already only be explained within the excitonic insulator scenario\,\cite{Cercellier2007}. 
In contrast, we propose that the gap component that cannot be quenched is a signature of an order parameter with short coherence length, which transfers into a pseudogap state with short range CDW fluctuations after optical excitation and that exhibits mostly Jahn-Teller character. The short correlation length associated with this pseudogap phase is in agreement with the reported loss of coherence after optical excitation in a recent  ultrafast electron diffraction\,(UED) experiment\,\cite{Cheng2022}, and  is confirmed by multiple studies showing that doping with Pt, S or Cu can lead to the formation of local domains with persisting gap and CDW charge modulation, while the backfolded Se$_{4p}$* band requiring long range order is almost completely suppressed\,\cite{Mottas2019a, Zhao2007, Yan2017, Novello2017, Lee2021, Iavarone2012}. The Jahn-Teller scenario is further corroborated by ultrafast THz experiments reporting a signatures of a residual gap and a periodic lattice distortion even after the excitonic order was completely destroyed, caused by a remaining Jahn-Teller-like CDW\,\cite{Porer2014a}.  Moreover, the strong coupling of the lattice to the electronic structure is confirmed by oscillations of the Se$_{4p}$ valence band position at the frequency of the characteristic A1g phonon mode\,\cite{Duan2021}. The sum of these observations strongly hint towards Jahn-Teller like CDW fluctuations as the reason for the robustness of the gap, however it should be noted that also electron-hole fluctuations\,\cite{Monney2010a, Monney2012} cannot be completely ruled out. 
Finally we note that the melting of the CDW state likely occurs in a spatially incoherent way: A previous equilibrium ARPES study\,\cite{Chen2016a} found that  when decreasing temperature a 2D (2$\times$2) CDW order sets in $\sim$30\,K above the well known 3D (2$\times$2$\times$2) structure, indicating a dimensional crossover. Moreover a recent UED study\,\cite{Cheng2022} found that optical excitation above a critical fluence melts the 3D CDW phase while creating a nonequilibrium 2D CDW phase. The structural dynamics probed by UED show noteworthy parallels to the behavior of the gap in the electronic spectrum studied in this work and are a promising topic of further theoretical and experimental investigations. \\

In conclusion this work places \tise in the same context of other quantum materials where a pseudogap phase appears to precede long-range order. We find strong indications that the gap in \tise is governed by two contributions, one caused by an excitonic condensate with long coherence length and one of Jahn-Teller character with short coherence length. The latter contribution can still give a well defined gap even after long range order was destroyed. In contrast, the well studied intensity of the folded Se$_{4p}$* band gives access to predominantly only one of these contributions and originates mostly from the excitonic order. Thus, in summary our work exemplifies how valuable mechanistic insight can be gained by studying the order parameter and  shines new light onto the complicated interplay of long and short range order in \ttise, which is eventually the key for the understanding of doping or pressure induced superconductivity in this and other materials. 

\section*{Data Availability} 
The datasets generated and/or analysed during the current study are not publicly available as they contain additional findings not reported in this manuscript,  but are available from the corresponding author on reasonable request.




\section*{Refrerences}
\bibliographystyle{IEEEtran}
\bibliography{main_paper_final2}

\begin{thebibliography}{10}
\providecommand{\url}[1]{#1}
\csname url@samestyle\endcsname
\providecommand{\newblock}{\relax}
\providecommand{\bibinfo}[2]{#2}
\providecommand{\BIBentrySTDinterwordspacing}{\spaceskip=0pt\relax}
\providecommand{\BIBentryALTinterwordstretchfactor}{4}
\providecommand{\BIBentryALTinterwordspacing}{\spaceskip=\fontdimen2\font plus
\BIBentryALTinterwordstretchfactor\fontdimen3\font minus
  \fontdimen4\font\relax}
\providecommand{\BIBforeignlanguage}[2]{{%
\expandafter\ifx\csname l@#1\endcsname\relax
\typeout{** WARNING: IEEEtran.bst: No hyphenation pattern has been}%
\typeout{** loaded for the language `#1'. Using the pattern for}%
\typeout{** the default language instead.}%
\else
\language=\csname l@#1\endcsname
\fi
#2}}
\providecommand{\BIBdecl}{\relax}
\BIBdecl

\bibitem{Gruener}
G.~Gr{\"{u}}ner, \emph{{Density Waves in Solids}}, 1st~ed., {L.P. Gor'kov G.
  Gr{\"{u}}ner}, Ed.\hskip 1em plus 0.5em minus 0.4em\relax Cambridge: Perseus
  Publishing, 1994.

\bibitem{Rossnagel2011}
K.~Rossnagel, ``{On the origin of charge-density waves in select layered
  transition-metal dichalcogenides},'' \emph{Journal of Physics Condensed
  Matter}, vol.~23, no.~21, 2011.

\bibitem{McMillan}
W.~L. McMillan, ``{Microscopic model of charge-density waves in 2H-TaSe$_2$},''
  \emph{Physical Review B}, vol.~16, no.~2, pp. 643--650, jul 1977.

\bibitem{Hellgren2017}
M.~Hellgren, J.~Baima, R.~Bianco, M.~Calandra, F.~Mauri, and L.~Wirtz,
  ``{Critical Role of the Exchange Interaction for the Electronic Structure and
  Charge-Density-Wave Formation in TiSe$_2$},'' \emph{Physical Review Letters},
  vol. 119, no.~17, pp. 1--6, 2017.

\bibitem{Lian2020}
C.~Lian, S.~J. Zhang, S.~Q. Hu, M.~X. Guan, and S.~Meng, ``{Ultrafast charge
  ordering by self-amplified exciton–phonon dynamics in TiSe$_2$},''
  \emph{Nature Communications}, vol.~11, no.~1, 2020.

\bibitem{Bianco2015}
R.~Bianco, M.~Calandra, and F.~Mauri, ``{Electronic and vibrational properties
  of TiSe$_2$ in the charge-density-wave phase from first principles},''
  \emph{Physical Review B - Condensed Matter and Materials Physics}, vol.~92,
  no.~9, pp. 1--19, 2015.

\bibitem{Kanigel2008}
A.~Kanigel, U.~Chatterjee, M.~Randeria, M.~R. Norman, G.~Koren, K.~Kadowaki,
  and J.~C. Campuzano, ``{Evidence for pairing above the transition temperature
  of cuprate superconductors from the electronic dispersion in the pseudogap
  phase},'' \emph{Physical Review Letters}, vol. 101, no.~13, pp. 1--4, 2008.

\bibitem{Ding1996}
H.~Ding, T.~Yokoya, J.~C. Campuzano, T.~Takahashi, M.~Randeria, M.~R. Norman,
  T.~Mochiku, K.~Kadowaki, and J.~Giapintzakis, ``{Spectroscopic evidence for a
  pseudogap in the normal state of underdoped high-T(c) superconductors},''
  \emph{Nature}, vol. 382, no. 6586, pp. 51--54, 1996.

\bibitem{Kogar2017}
A.~Kogar, M.~S. Rak, S.~Vig, A.~A. Husain, F.~Flicker, Y.~I. Joe, L.~Venema,
  G.~J. Macdougall, T.~C. Chiang, E.~Fradkin, J.~V. Wezel, and P.~Abbamonte,
  ``{Signatures of exciton condensation in a transition metal
  dichalcogenide},'' \emph{Science}, vol. 1317, no. December, pp. 1314--1317,
  2017.

\bibitem{Monney2009}
C.~Monney, H.~Cercellier, F.~Clerc, C.~Battaglia, E.~F. Schwier, C.~Didiot,
  M.~G. Garnier, H.~Beck, P.~Aebi, H.~Berger, L.~Forr{\'{o}}, and L.~Patthey,
  ``{Spontaneous exciton condensation in 1T$-$TiSe$_2$: BCS-like approach},''
  \emph{Physical Review B - Condensed Matter and Materials Physics}, vol.~79,
  no.~4, pp. 1--11, 2009.

\bibitem{Weber2011}
F.~Weber, S.~Rosenkranz, J.~P. Castellan, R.~Osborn, G.~Karapetrov, R.~Hott,
  R.~Heid, K.~P. Bohnen, and A.~Alatas, ``{Electron-phonon coupling and the
  soft phonon mode in TiSe$_2$},'' \emph{Physical Review Letters}, vol. 107,
  no.~26, pp. 1--5, 2011.

\bibitem{Otto2021}
M.~R. Otto, J.-H. Pöhls, L.~P. René~de Cotret, M.~J. Stern, M.~Sutton, and
  B.~J. Siwick, ``{Mechanisms of electron-phonon coupling unraveled in momentum
  and time: The case of soft phonons in TiSe$_2$},'' \emph{Nature
  Communications}, vol.~7, no.~20, 2021.

\bibitem{Kusmartseva2009}
A.~F. Kusmartseva, B.~Sipos, H.~Berger, L.~Forr{\'{o}}, and E.~Tuti{\v{s}},
  ``{Pressure Induced Superconductivity in Pristine 1T$-$TiSe$_2$},''
  \emph{Physical Review Letters}, vol. 103, no.~23, pp. 2--5, 2009.

\bibitem{Joe2014}
Y.~I. Joe, X.~M. Chen, P.~Ghaemi, K.~D. Finkelstein, G.~A. {De La Pe{\~{n}}a},
  Y.~Gan, J.~C. Lee, S.~Yuan, J.~Geck, G.~J. MacDougall, T.~C. Chiang, S.~L.
  Cooper, E.~Fradkin, and P.~Abbamonte, ``{Emergence of charge density wave
  domain walls above the superconducting dome in 1T$-$TiSe$_2$},'' \emph{Nature
  Physics}, vol.~10, no.~6, pp. 421--425, 2014.

\bibitem{Ramirez2006}
A.~P. Ramirez, N.~P. Ong, and R.~J. Cava, ``{Superconductivity in
  Cu$_x$TiSe$_2$},'' \emph{Nature Physics}, vol.~2, no. August, pp. 18--22,
  2006.

\bibitem{Hildebrand2016}
B.~Hildebrand, T.~Jaouen, C.~Didiot, E.~Razzoli, G.~Monney, M.~L. Mottas,
  A.~Ubaldini, H.~Berger, C.~Barreteau, H.~Beck, D.~R. Bowler, and P.~Aebi,
  ``{Short-range phase coherence and origin of the 1T$-$TiSe$_2$ charge density
  wave},'' \emph{Physical Review B}, vol.~93, no.~12, pp. 1--5, 2016.

\bibitem{Porer2014a}
M.~Porer, U.~Leierseder, J.~M. M{\'{e}}nard, H.~Dachraoui, L.~Mouchliadis,
  I.~E. Perakis, U.~Heinzmann, J.~Demsar, K.~Rossnagel, and R.~Huber,
  ``{Non-thermal separation of electronic and structural orders in a persisting
  charge density wave},'' \emph{Nature Materials}, vol.~13, no.~9, pp.
  857--861, 2014.

\bibitem{Hedayat2019a}
H.~Hedayat, C.~J. Sayers, D.~Bugini, C.~Dallera, D.~Wolverson, T.~Batten,
  S.~Karbassi, S.~Friedemann, G.~Cerullo, J.~van Wezel, S.~R. Clark,
  E.~Carpene, and E.~{Da Como}, ``{Excitonic and lattice contributions to the
  charge density wave in 1T-TiSe$_2$ revealed by a phonon bottleneck},''
  \emph{Physical Review Research}, vol.~1, no.~2, pp. 1--11, 2019.

\bibitem{Duan2021}
S.~Duan, Y.~Cheng, W.~Xia, Y.~Yang, C.~Xu, F.~Qi, C.~Huang, T.~Tang, Y.~Guo,
  W.~Luo, D.~Qian, D.~Xiang, J.~Zhang, and W.~Zhang, ``{Optical manipulation of
  electronic dimensionality in a quantum material},'' \emph{Nature}, vol. 595,
  no. 7866, pp. 239--244, 2021.

\bibitem{Buss2019}
J.~H. Buss, H.~Wang, Y.~Xu, C.~Jozwiak, J.~Pepper, J.~Maklar, F.~Joucken,
  L.~Zeng, S.~Stoll, Z.~Hussain, A.~Lanzara, Y.-d. Chuang, J.~D. Denlinger,
  H.~Wang, Y.~Xu, J.~Maklar, L.~Zeng, J.~Pepper, and Y.-d. Chuang, ``{A setup
  for extreme-ultraviolet ultrafast angle-resolved photoelectron spectroscopy
  at 50-kHz repetition rate},'' \emph{Review of Scientific Instruments},
  vol.~90, no. 023105, 2019.

\bibitem{Wang2015}
H.~Wang, Y.~Xu, S.~Ulonska, J.~S. Robinson, P.~Ranitovic, and R.~A. Kaindl,
  ``{Bright high-repetition-rate source of narrowband extreme-ultraviolet
  harmonics beyond 22 eV},'' \emph{Nature Communications}, vol.~6, no. 7459,
  2015.

\bibitem{Watson2019}
M.~D. Watson, O.~J. Clark, F.~Mazzola, I.~Markovi{\'{c}}, V.~Sunko, T.~K. Kim,
  K.~Rossnagel, and P.~D. King, ``{Orbital- and kz -Selective Hybridization of
  Se 4p and Ti 3d States in the Charge Density Wave Phase of TiSe$_2$},''
  \emph{Physical Review Letters}, vol. 122, no.~7, pp. 1--6, 2019.

\bibitem{Mathias2016}
S.~Mathias, S.~Eich, J.~Urbancic, S.~Michael, A.~V. Carr, S.~Emmerich,
  A.~Stange, T.~Popmintchev, T.~Rohwer, M.~Wiesenmayer, A.~Ruffing, S.~Jakobs,
  S.~Hellmann, P.~Matyba, C.~Chen, L.~Kipp, M.~Bauer, H.~C. Kapteyn, H.~C.
  Schneider, K.~Rossnagel, M.~M. Murnane, and M.~Aeschlimann, ``{Self-amplified
  photo-induced gap quenching in a correlated electron material},''
  \emph{Nature Communications}, vol.~7, pp. 1--8, 2016.

\bibitem{Monney2016}
C.~Monney, M.~Puppin, C.~W. Nicholson, M.~Hoesch, R.~T. Chapman, E.~Springate,
  H.~Berger, A.~Magrez, C.~Cacho, R.~Ernstorfer, and M.~Wolf, ``{Revealing the
  role of electrons and phonons in the ultrafast recovery of charge density
  wave correlations in 1T$-$ TiSe$_2$},'' \emph{Physical Review B}, vol.~94,
  no.~16, pp. 1--9, 2016.

\bibitem{Monney2010a}
C.~Monney, E.~F. Schwier, M.~G. Garnier, N.~Mariotti, C.~Didiot, H.~Beck,
  P.~Aebi, H.~Cercellier, J.~Marcus, C.~Battaglia, H.~Berger, and A.~N. Titov,
  ``{Temperature-dependent photoemission on 1T -TiSe$_2$: Interpretation within
  the exciton condensate phase model},'' \emph{Physical Review B - Condensed
  Matter and Materials Physics}, vol.~81, no.~15, pp. 1--9, 2010.

\bibitem{DiSalvo1976}
F.~J. {Di Salvo}, D.~E. Moncton, and J.~V. Waszczak, ``{Electronic properties
  and superlattice formation in the semimetal TiSe$_2$},'' \emph{Physical
  Review B}, vol.~14, no.~10, pp. 4321--4328, nov 1976.

\bibitem{Rossnagel2002}
K.~Rossnagel, L.~Kipp, and M.~Skibowski, ``{Charge-density-wave phase
  transition in (formula presented): Excitonic insulator versus band-type
  Jahn-Teller mechanism},'' \emph{Physical Review B - Condensed Matter and
  Materials Physics}, vol.~65, no.~23, pp. 1--7, 2002.

\bibitem{Rossnagel2010}
K.~Rossnagel, ``{Suppression and emergence of charge-density waves at the
  surfaces of layered 1T$-$TiSe$_2$ and 1T$-$TaS$_2$ by in situ Rb
  deposition},'' \emph{New Journal of Physics}, vol.~12, 2010.

\bibitem{Rohwer2011}
T.~Rohwer, S.~Hellmann, M.~Wiesenmayer, C.~Sohrt, A.~Stange, B.~Slomski,
  A.~Carr, Y.~Liu, L.~M. Avila, M.~Kall{\"{a}}signne, S.~Mathias, L.~Kipp,
  K.~Rossnagel, and M.~Bauer, ``{Collapse of long-range charge order tracked by
  time-resolved photoemission at high momenta},'' \emph{Nature}, vol. 471, no.
  7339, pp. 490--494, 2011.

\bibitem{Kidd2002}
T.~E. Kidd, T.~Miller, M.~Y. Chou, and T.~C. Chiang, ``{Electron-hole coupling
  and the charge density wave transition in TiSe$_2$},'' \emph{Physical Review
  Letters}, vol.~88, no.~22, pp. 226\,402/1--226\,402/4, 2002.

\bibitem{Hellmann2012}
S.~Hellmann, T.~Rohwer, M.~Kall{\"{a}}ne, K.~Hanff, C.~Sohrt, A.~Stange,
  A.~Carr, M.~M. Murnane, H.~C. Kapteyn, L.~Kipp, M.~Bauer, and K.~Rossnagel,
  ``{Time-domain classification of charge-density-wave insulators},''
  \emph{Nature Communications}, vol.~3, 2012.

\bibitem{Rohde$_2$013}
G.~Rohde, T.~Rohwer, C.~Sohrt, A.~Stange, S.~Hellmann, L.~X. Yang, K.~Hanff,
  A.~Carr, M.~M. Murnane, H.~Kapteyn, L.~Kipp, K.~Rossnagel, and M.~Bauer,
  ``{Tracking the relaxation pathway of photo-excited electrons in
  1T$-$TiSe$_2$},'' \emph{European Physical Journal: Special Topics}, vol. 222,
  no.~5, pp. 997--1004, 2013.

\bibitem{Rohde$_2$014}
G.~Rohde, T.~Rohwer, A.~Stange, C.~Sohrt, K.~Hanff, L.~X. Yang, L.~Kipp,
  K.~Rossnagel, and M.~Bauer, ``{Does the excitation wavelength affect the
  ultrafast quenching dynamics of the charge-density wave in 1T$-$TiSe$_2$?}''
  \emph{Journal of Electron Spectroscopy and Related Phenomena}, vol. 195, pp.
  244--248, 2014.

\bibitem{Burian}
M.~Burian, M.~Porer, J.~R.~L. Mardegan, V.~Esposito, S.~Parchenko, B.~Burganov,
  N.~Gurung, M.~Ramakrishnan, V.~Scagnoli, H.~Ueda, S.~Francoual, F.~Fabrizi,
  Y.~Tanaka, T.~Togashi, Y.~Kubota, M.~Yabashi, K.~Rossnagel, S.~L. Johnson,
  and U.~Staub, ``{Structural involvement in the melting of the charge density
  wave in 1T$-$TiSe$_2$},'' \emph{Physical Review Research}, vol.~3, no.~1, p.
  13128, 2021.

\bibitem{Mohr-Vorobeva2011}
E.~M{\"{o}}hr-Vorobeva, S.~L. Johnson, P.~Beaud, U.~Staub, R.~{De Souza},
  C.~Milne, G.~Ingold, J.~Demsar, H.~Schaefer, and A.~Titov, ``{Nonthermal
  melting of a charge density wave in TiSe$_2$},'' \emph{Physical Review
  Letters}, vol. 107, no.~3, pp. 1--4, 2011.

\bibitem{Cheng2022}
Y.~Cheng, A.~Zong, J.~Li, W.~Xia, S.~Duan, W.~Zhao, Y.~Li, F.~Qi, J.~Wu,
  L.~Zhao, P.~Zhu, X.~Zou, T.~Jiang, Y.~Guo, L.~Yang, D.~Qian, W.~Zhang,
  A.~Kogar, M.~W. Zuerch, D.~Xiang, and J.~Zhang, ``{Light-induced dimension
  crossover dictated by excitonic correlations},'' \emph{Nature
  Communications}, vol.~13, no. 963, 2022.

\bibitem{Singh2017}
B.~Singh, C.~H. Hsu, W.~F. Tsai, V.~M. Pereira, and H.~Lin, ``{Stable charge
  density wave phase in a 1T$-$TiSe$_2$ monolayer},'' \emph{Physical Review B},
  vol.~95, no.~24, pp. 1--7, 2017.

\bibitem{Smallwood2014}
C.~L. Smallwood, W.~Zhang, T.~L. Miller, C.~Jozwiak, H.~Eisaki, D.~H. Lee, and
  A.~Lanzara, ``{Time- and momentum-resolved gap dynamics in
  Bi$_2$Sr$_2$CaCu$_2$O$_{8+\delta}$},'' \emph{Physical Review B - Condensed
  Matter and Materials Physics}, vol.~89, no.~11, pp. 1--8, 2014.

\bibitem{Schmitt2008}
F.~Schmitt, P.~S. Kirchmann, U.~Bovensiepen, R.~G. Moore, L.~Rettig, M.~Krenz,
  J.~H. Chu, N.~Ru, L.~Perfetti, D.~H. Lu, M.~Wolf, I.~R. Fisher, and Z.~X.
  Shen, ``{Transient electronic structure and melting of a charge density wave
  in TbTe$_3$},'' \emph{Science}, vol. 321, no. 5896, pp. 1649--1652, 2008.

\bibitem{Perfetti2006}
L.~Perfetti, P.~A. Loukakos, M.~Lisowski, U.~Bovensiepen, H.~Berger,
  S.~Biermann, P.~S. Cornaglia, A.~Georges, and M.~Wolf, ``{Time evolution of
  the electronic structure of 1T$-$TaS$_2$ through the insulator-metal
  transition},'' \emph{Physical Review Letters}, vol.~97, no.~6, pp. 1--4,
  2006.

\bibitem{Petersen2011}
J.~C. Petersen, S.~Kaiser, N.~Dean, A.~Simoncig, H.~Y. Liu, A.~L. Cavalieri,
  C.~Cacho, I.~C. Turcu, E.~Springate, F.~Frassetto, L.~Poletto, S.~S. Dhesi,
  H.~Berger, and A.~Cavalleri, ``{Clocking the melting transition of charge and
  lattice order in 1T$-$TaS$_2$ with ultrafast extreme-ultraviolet
  angle-resolved photoemission spectroscopy},'' \emph{Physical Review Letters},
  vol. 107, no.~17, pp. 1--5, 2011.

\bibitem{Zhang2014}
W.~Zhang, C.~Hwang, C.~L. Smallwood, T.~L. Miller, G.~Affeldt, K.~Kurashima,
  C.~Jozwiak, H.~Eisaki, T.~Adachi, Y.~Koike, D.~H. Lee, and A.~Lanzara,
  ``{Ultrafast quenching of electron-boson interaction and superconducting gap
  in a cuprate superconductor},'' \emph{Nature Communications}, vol.~5, pp.
  1--6, 2014.

\bibitem{Smallwood2015}
C.~L. Smallwood, W.~Zhang, T.~L. Miller, G.~Affeldt, K.~Kurashima, C.~Jozwiak,
  T.~Noji, Y.~Koike, H.~Eisaki, D.~H. Lee, R.~A. Kaindl, and A.~Lanzara,
  ``{Influence of optically quenched superconductivity on quasiparticle
  relaxation rates in Bi$_2$Sr$_2$CaCu$_2$O$_{8+\delta}$},'' \emph{Physical
  Review B - Condensed Matter and Materials Physics}, vol.~92, no.~16, pp.
  1--6, 2015.

\bibitem{Zhang2016}
W.~Zhang, T.~Miller, C.~L. Smallwood, Y.~Yoshida, H.~Eisaki, R.~A. Kaindl,
  D.~H. Lee, and A.~Lanzara, ``{Stimulated emission of Cooper pairs in a
  high-temperature cuprate superconductor},'' \emph{Scientific Reports},
  vol.~6, pp. 1--7, 2016.

\bibitem{Zong2019}
A.~Zong, A.~Kogar, Y.~Q. Bie, T.~Rohwer, C.~Lee, E.~Baldini, E.~Erge{\c{c}}en,
  M.~B. Yilmaz, B.~Freelon, E.~J. Sie, H.~Zhou, J.~Straquadine, P.~Walmsley,
  P.~E. Dolgirev, A.~V. Rozhkov, I.~R. Fisher, P.~Jarillo-Herrero, B.~V. Fine,
  and N.~Gedik, ``{Evidence for topological defects in a photoinduced phase
  transition},'' \emph{Nature Physics}, vol.~15, no.~1, pp. 27--31, 2019.

\bibitem{Saha2021}
T.~Saha, D.~Gole{\v{z}}, G.~{De Ninno}, J.~Mravlje, Y.~Murakami, B.~Ressel,
  M.~Stupar, and P.~R. Ribi{\v{c}}, ``{Photoinduced phase transition and
  associated timescales in the excitonic insulator Ta$_2$NiSe$_5$},''
  \emph{Physical Review B}, vol. 103, no.~14, pp. 1--13, 2021.

\bibitem{GoleZ2016}
D.~Gole{\v{z}}, P.~Werner, and M.~Eckstein, ``{Photoinduced gap closure in an
  excitonic insulator},'' \emph{Physical Review B}, vol.~94, no.~3, pp. 1--11,
  2016.

\bibitem{Cercellier2007}
H.~Cercellier, C.~Monney, F.~Clerc, C.~Battaglia, L.~Despont, M.~G. Garnier,
  H.~Beck, P.~Aebi, L.~Patthey, H.~Berger, and L.~Forr{\'{o}}, ``{Evidence for
  an excitonic insulator phase in 1T$-$TiSe$_2$},'' \emph{Physical Review
  Letters}, vol.~99, no.~14, pp. 1--4, 2007.

\bibitem{Mottas2019a}
M.~L. Mottas, T.~Jaouen, B.~Hildebrand, M.~Rumo, F.~Vanini, E.~Razzoli,
  E.~Giannini, C.~Barreteau, D.~R. Bowler, C.~Monney, H.~Beck, and P.~Aebi,
  ``{Semimetal-to-semiconductor transition and charge-density-wave suppression
  in 1T$-$TiSe$_{2-x}$S$_x$ single crystals},'' \emph{Physical Review B},
  vol.~99, no.~15, p. 155103, 2019.

\bibitem{Zhao2007}
J.~F. Zhao, H.~W. Ou, G.~Wu, B.~P. Xie, Y.~Zhang, D.~W. Shen, J.~Wei, L.~X.
  Yang, J.~K. Dong, M.~Arita, H.~Namatame, M.~Taniguchi, X.~H. Chen, and D.~L.
  Feng, ``{Evolution of the Electronic Structure of 1T$-$Cu$_x$TiSe$_2$},''
  \emph{Physical Review Letters}, vol. 146401, no.~99, pp. 2--5, 2007.

\bibitem{Yan2017}
S.~Yan, D.~Iaia, E.~Morosan, E.~Fradkin, P.~Abbamonte, and V.~Madhavan,
  ``{Influence of Domain Walls in the Incommensurate Charge Density Wave State
  of Cu Intercalated 1T$-$TiSe$_2$},'' \emph{Physical Review Letters}, vol.
  118, no.~10, pp. 1--5, 2017.

\bibitem{Novello2017}
A.~M. Novello, M.~Spera, A.~Scarfato, A.~Ubaldini, E.~Giannini, D.~R. Bowler,
  and C.~Renner, ``{Stripe and Short Range Order in the Charge Density Wave of
  1T$-$Cu$_x$TiSe$_2$},'' \emph{Physical Review Letters}, vol. 118, no.~1, pp.
  1--5, 2017.

\bibitem{Lee2021}
K.~Lee, J.~Choe, D.~Iaia, J.~Li, J.~Zhao, M.~Shi, J.~Ma, M.~Yao, Z.~Wang, C.~L.
  Huang, M.~Ochi, R.~Arita, U.~Chatterjee, E.~Morosan, V.~Madhavan, and
  N.~Trivedi, ``{Metal-to-insulator transition in Pt-doped TiSe$_2$ driven by
  emergent network of narrow transport channels},'' \emph{npj Quantum
  Materials}, vol.~6, no.~1, 2021.

\bibitem{Iavarone2012}
M.~Iavarone, R.~{Di Capua}, X.~Zhang, M.~Golalikhani, S.~A. Moore, and
  G.~Karapetrov, ``{Evolution of the charge density wave state in
  Cu$_x$TiSe$_2$},'' \emph{Physical Review B - Condensed Matter and Materials
  Physics}, vol.~85, no.~15, pp. 1--6, 2012.

\bibitem{Monney2012}
C.~Monney, G.~Monney, P.~Aebi, and H.~Beck, ``{Electron-hole fluctuation phase
  in 1T$-$TiSe$_2$},'' \emph{Physical Review B - Condensed Matter and Materials
  Physics}, vol.~85, no.~23, pp. 1--8, 2012.

\bibitem{Chen2016a}
P.~Chen, Y.~Chan, X.~Fang, S.~Mo, Z.~Hussain, A.~Fedorov, M.~Chou, and
  T.~Chiang, ``{Hidden Order and Dimensional Crossover of the Charge Density
  Waves in TiSe$_2$},'' \emph{Scientific Reports}, vol.~6, no. 37910, 2016.

\end{thebibliography}

\section*{Acknowledgments}
This work was primarily funded by the U.S. Department of Energy (DOE), Office of Science, Office of Basic Energy Sciences, Materials Sciences and Engineering Division under contract no. DE-AC02-05CH11231 (Ultrafast Materials Science program KC2203). S.T acknowledges support from NSF CMMI 1933214 and DMR 2129412 for material development and NSF ECCS 2052547 and DMR 2111812 for structure-performance relations.

\section*{Competing Interest}
The authors declare no competing financial interest.

\section*{Author Contributions}
A.L.\ designed and supervised the project. Data was collected by M.H.\ and Y.L.\ and analyzed by M.H.\ with help from Y.L., N.D., R.K.\ and A.L.\ Samples were provided by R.S.\ and S.T.\ The XUV-trARPES setup was designed by R.K.\ Manuscript preparation was done by M.H.\ with input from all co-authors.  


\end{document}